\documentclass[preprint2]{aastex}
\usepackage{psfig}

\newcommand\pp     {$\pm$}

\newcommand\nburst {$\nu_{burst}$}
\newcommand\nspin {$\nu_{spin}$}
\newcommand\norbit {$\nu_{orbit}$}
\newcommand\Dn   {$\Delta\nu$}

\righthead{Quasi-periodic X-ray brightness fluctuations in an accreting 
millisecond pulsar}

\slugcomment{}

\begin{document}

\title{Quasi-periodic X-ray brightness fluctuations in an accreting 
millisecond pulsar}

\author{Rudy Wijnands\altaffilmark{1}, Michiel van der Klis\altaffilmark{2},
 Jeroen Homan\altaffilmark{3}, Deepto Chakrabarty\altaffilmark{4},
 Craig B. Markwardt\altaffilmark{5}, Ed H. Morgan\altaffilmark{4}}

\altaffiltext{1}{School of Physics and Astronomy,
University of St Andrews, North Haugh, St Andrews, Fife KY16 9SS,
Scotland, UK; radw@st-andrews.ac.uk}

\altaffiltext{2}{Astronomical Institute 'Anton Pannekoek', University of 
Amsterdam, and Centre for High-Energy Astrophysics, Kruislaan 403,
1098 SJ, Amsterdam, The Netherlands}

\altaffiltext{3}{INAF-Osservatorio Astronomico di Brera, Via E. Bianchi 46,
23807 Merate LC, Italy}

\altaffiltext{4}{Center for Space Research, Massachusetts Institute of
Technology, 77 Massachusetts Avenue, Cambridge, Massachusetts 02139,
USA}

\altaffiltext{5}{Laboratory for High Energy Astrophysics, NASA Goddard Space Flight Center, Greenbelt, Maryland 20771, USA}

\begin{abstract}
{\bf The relativistic plasma flows onto neutron stars that are
accreting material from stellar companions can be used to probe
strong-field gravity as well as the physical conditions in the
supranuclear-density interiors of neutron stars. Plasma
inhomogeneities orbiting a few kilometres above the stars are
observable as X-ray brightness fluctuations on the millisecond
dynamical timescale of the flows$^{1-3}$. Two frequencies in the
kilohertz range dominate these fluctuations: the twin kilohertz
quasi-periodic oscillations (kHz QPOs). Competing models for the
origins of these oscillations (based on orbital motions) all predict
that they should be related to the stellar spin frequency$^{4-10}$,
but tests have been difficult because the spins were not unambiguously
known. Here we report the detection of kHz QPOs from a pulsar whose
spin frequency is known. Our measurements establish a clear link
between kHz QPOs and stellar spin, but one not predicted by any
current model. A new approach to understanding kHz QPOs is now
required. We suggest that a resonance between the spin and general
relativistic orbital and epicyclic frequencies could provide the
observed relation between QPOs and spin.}\\

\end{abstract}

Twin kHz QPOs with frequencies between 300 and 1,300 Hz occur in more
than 20 accreting neutron stars$^{3}$. In seven of these, during
thermonuclear X-ray bursts caused by the explosion of material
accumulated on the neutron-star surface, short-lived oscillations are
observed with frequencies \nburst~in the range 270--620 Hz,
approximately constant but different in each source, that are thought
to be caused by the neutron-star spin$^{11}$. The frequency difference
\Dn~between the two kHz QPOs is close to either \nburst, or \nburst/2,
depending on the source. This observed commensurability has given rise
to beat-frequency models$^{6,12-14}$. Such models identify the
higher-frequency (``upper'') kHz QPO with the frequency of orbital
motion, closely around the neutron star, of the plasma at the inner
edge of the accretion disk (within this radius no stable orbits exist
and the matter plunges in), and the other (``lower'') kHz QPO with a
rotational beat interaction between the upper kHz QPO and the
neutron-star spin. These models predict \Dn~to be equal to the spin
frequency \nspin. In this interpretation, we have either \nburst~=
\nspin~or \nburst~= 2\nspin, depending on whether one or two hotspots
form on the star's surface during the bursts. Testing this requires
searching for harmonic structure in the burst oscillations, but
searches$^{15}$ are limited in terms of statistics as compared to the
case of a true pulsar.

A problem for beat-frequency models is the fact that \Dn~is neither
constant$^{16,17}$, nor exactly equal$^{18,19}$ to \nspin~(but
possible solutions in the beat-frequency models do exist$^{20}$). This
has motivated the proposal of several alternative models to explain
the kHz QPOs$^{7-10}$, all of which, like beat-frequency models, use
orbital motion as one of the observed frequencies, but which use
another mechanism -- such as general relativistic epicyclic motion,
which is very fast in these extreme gravitational fields -- to
generate the other frequency. None of these models predict
commensurability between spin and kHz QPO frequencies such as
beat-frequency models do. It has been clear for some time that
independent measurements of the neutron-star spin frequency could
provide the definitive test of the beat-frequency idea. The
discovery$^{21}$ of the first accreting millisecond pulsar in 1998
raised the prospect of such a test, but until now no accreting
millisecond pulsar has shown the anticipated kHz QPOs.

On 13 October 2002, a bright new outburst of the recurrent X-ray
transient and accreting millisecond pulsar, SAX J1808.4--3658,
started$^{22}$. Using the Rossi X-ray Timing Explorer satellite, we
obtained $\sim$700 ks of data covering the outburst with daily
observations from 15 October to 26 November 2002. We performed a
series of fast Fourier transforms to search for kHz QPOs, and
discovered several. One kHz QPO was monitored for 12 days
(Fig. 1). Its frequency increased from 570 to 725 Hz between 15 and 16
October, then gradually decreased to a minimum of 280 Hz on 21
October, suddenly increased again to 400 Hz the next day, and then
resumed its decrease, to 320 Hz on 26 October. After this the QPO was
not detected, presumably owing to the limited statistics resulting
from the drop in count rate (by a factor of about 4) by this stage of
the outburst. On 16 October, a second kHz QPO was detected
simultaneously at a frequency of \Dn~= 195\pp 6Hz below the first one
(Fig. 1, top panel). In the remaining data, this second QPO was not
detected with a significance greater than 2.5$\sigma$. Additional
broad noise structures of marginal significance are sometimes present
around 350, 220 and 120 Hz.

The frequencies, strengths and coherences of these two kHz QPOs in SAX
J1808.4--3658, as well as the variations in their parameters as a
function of source luminosity, are very similar to those of the twin
kHz QPOs observed in those neutron-star low mass binaries for which no
pulsations are seen in their persistent emission (the non-pulsing
sources$^3$), so we interpret this as the same phenomenon. The
$\sim$195-Hz frequency difference between the two QPOs is the lowest
known (in other systems \Dn~ranges between 225 and 350 Hz)$^3$. It is
far below the 401-Hz spin frequency predicted in simple beat-frequency
models$^{6,12}$, yet there is a clear commensurability: \Dn~is
consistent with \nspin/2. This is reminiscent of those non-pulsing
sources where \Dn~is near \nburst/2. Hence our results support the
spin interpretation of the burst oscillations in those sources, as
well as the suspicion that \nburst~is always near \nspin~(and never
twice that). Moreover, as reported in a companion paper$^{23}$, in our
observations we also detected four thermonuclear bursts in which we
discovered burst oscillations at the neutron-star spin frequency. This
discovery further strengthens the idea that the burst oscillations are
always at the stellar spin frequency.

We now demonstrate that, unless we are wrong about the spin frequency
of this pulsar, which seems highly unlikely, our observations falsify
current beat-frequency models, yet so pose a severe challenge to all
other current kHz QPO models. In beat-frequency models$^{5,6,24}$, the
spin-orbit beat interaction occurs at a frequency \hbox{$\nu_{beat} =
n$(\norbit -- \nspin)}, where \norbit~is the orbital frequency, and
the positive integer $n$ is a symmetry factor accounting for multiple
symmetrically located spin-orbit interaction sites (for example, $n =
2$ for two magnetic poles). Simple beat-frequency models have $n = 1$
and the two kHz QPOs at $\nu_{beat}$ and \norbit, so that \hbox{\Dn~=
\norbit -- $\nu_{beat}$ =
\nspin}. As noted, this is clearly inconsistent with our
observations. In general, $n$ is the least common multiple (LCM) of
the number of stellar and the number of orbital interaction sites: $n
=$ LCM($n_{spin}$, $n_{orbit}$). Whatever the value of these numbers,
and even allowing the observed orbital QPO to be at either \norbit~or
$n_{orbit}$\norbit, there is no configuration predicting the observed
\Dn~=
\nspin/2.

The only solution consistent with current beat-frequency models is
that \nspin~is half the observed 401-Hz pulsar frequency (that is, two
hotspots) and $n = 1$ (that is, only one pulsar beam or magnetic pole
interacts with the orbit). However, the very stringent amplitude upper
limit of 0.014\% root mean square (r.m.s.; E.H.M. {\it et al.},
manuscript in preparation) on any coherent signal at 200.5 Hz (only
0.38\% of the signal at 401 Hz) makes this solution very unlikely: a
fine-tuned special geometry would be required to hide the true spin
frequency that well. We note that the sonic point beat-frequency
model$^6$ predicts a spectrum of many additional weaker QPO peaks
(higher-order sidebands), one of which (for example, abs($\nu_{upper}$
-- 3\nspin) or, a particularly close match, abs(3$\nu_{lower}$ --
2\nspin), where $\nu_{upper}$ and $\nu_{lower}$ are the upper and
lower kHz QPO frequency, respectively) could be picked out and
identified with a QPO that we observe. However, this would not explain
why, contrary to predictions, this peak is stronger rather than much
weaker than the main peaks. In our view, the coincidence between
\Dn~and \nburst/2 in several other sources strongly suggests instead
that the relation between kHz QPOs and neutron star spin is one where
(in SAX J1808.4--3658 and in those several other sources) the peaks
are separated by \nspin/2. This has not been predicted by any model.

Inspired by the relativistic resonance models of ref. 9 (see also refs
25--27 for related suggestions), we note that there are particular
radii in the disk where the difference between the general
relativistic orbital frequency ($\nu_\phi$) and radial frequency
($\nu_r$; note that $\nu_\phi-\nu_r$ is the periastron precession
frequency) is equal to
\nspin~and \nspin/2. For a 1.4-solar-mass Schwarzschild geometry,
these radii are $\sim$18 and $\sim$23 km, respectively (the best match
to the observed frequencies in fact occurs at 1.1 solar masses, but
see below). If a radiatively or magnetically mediated spin-orbit
interaction were to set up a resonance at one of those radii leading
to kHz QPOs at $\nu_\phi$ and $\nu_r$, this could explain the observed
frequency commensurabilities with no involvement of a beat
frequency. Whether
\Dn~is \nspin~or \nspin/2 might depend on whether or not the relevant
radius is within the inner edge of the disk; in support of this, we
note that all four systems where \Dn~$\approx$~\nspin/2 are those
where \nspin~exceeds 400 Hz, and vice versa for the remaining three
cases. That the frequencies can shift and the observed
commensurabilities are often not exact might be related to a mechanism
similar to that which makes the resonances occur off the naively
expected integer ratios in the numerical calculations of ref. 28. We
note that this combination of ideas (adding the spin as an ingredient
in the relativistic resonance model) can lead to a single framework
within which the apparently discrepant frequency commensurabilities
observed in black holes (which take the form of integer ratios) as
well as neutron stars can both be understood in terms of resonances at
privileged radii in an accretion disk, namely those radii where the
general relativistic epicyclic frequencies (and, in the case of the
neutron-star systems, spin) are commensurate.

A surprising third QPO (Fig. 2) near 410 Hz was detected on four
occasions, just above the 401-Hz pulse frequency. Its frequency
increased from 409.3\pp0.7 to 413.5\pp0.2 Hz while the frequency of
the upper kHz QPO increased from 328\pp4 to 390\pp5 Hz. This third kHz
QPO is a phenomenon not previously seen, and is probably related to
the pulsating nature of SAX J1808.4--3658. To some extent, it
resembles the sideband to the lower kHz QPO observed$^{29}$ in three
non-pulsing kHz QPO sources: the 410-Hz QPO might be a sideband to the
pulsation with a similar underlying mechanism. These lower kHz QPO
sidebands were suggested$^{29}$ to be due to Lense-Thirring
precession, a predicted -- but not yet observed -- general
relativistic wobble of the orbital plane (nodal precession). The
precession frequency is predicted$^{30}$ to be quadratically related
to the orbital frequency itself, and in the case of both the QPO
sidebands and the $\sim$410-Hz QPO this is consistent with
observations. However, as Lense-Thirring precession is prograde, a
beat between a pulsar beam and a precessing orbit would produce a
sideband below, rather than the observed sideband above, the pulsar
frequency. Another resonance, perhaps at the radius where the general
relativistic vertical epicyclic frequency matches \nspin, might be
considered as an explanation of this phenomenon.

\acknowledgments
We thank J. Swank and E. Smith for their efforts in scheduling our
RXTE observations of SAX J1808.4--3658, and D. Psaltis for
discussions. M.v.d.K. was supported by NWO.

\begin{figure}[t]
\begin{center}
\begin{tabular}{c}
\psfig{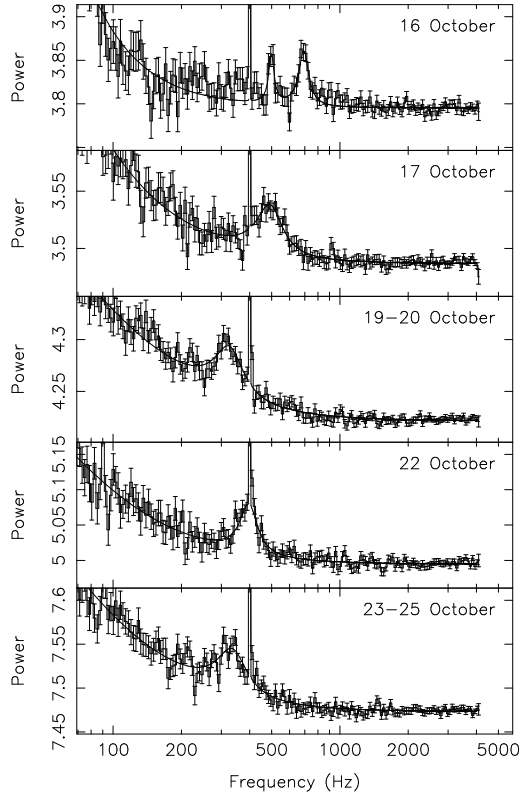}
\end{tabular}
\figcaption{\label{fig:f1}\small
The power-density spectra obtained for SAX J1808.4--3658 during its
2002 outburst. The power is given in units of (r.m.s./mean)$^2$/(Hz
$\times 10^{-3}$). These spectra show the presence of the pulsations
at 401 Hz, the QPOs, and broadband noise below a few hundred
hertz. The date the data were obtained is given in each panel. All
data during those days (for the energy range 3--60 keV) were used to
create the power-density spectra, except for that for 16 October for
which only the data between 0:53 and 10:00 UTC were used and for the
energy range 3.3--39 keV. The properties of the QPO were obtained by
fitting the power-density spectra with a constant for the Poisson
level (clearly visible at frequencies $>$1,000 Hz), and three
lorentzian functions to fit the continuum at low frequencies, the
pulsation frequency at 401 Hz, and the upper kHz QPO. On 16 October,
two kHz QPOs are visible with frequencies of 499\pp4 Hz and 694\pp4
Hz, fractional r.m.s. amplitudes (3.3--39 keV) of 5.2\pp0.5\%
(5$\sigma$) and 9.1\pp0.5\% (9$\sigma$) and widths (FWHM) of 35\pp12
Hz and 82\pp15 Hz, which we identify as the lower and the upper kHz
QPO, respectively. In the remaining observations one kHz QPO is
detected which we identify as the upper kHz QPO based on its factional
r.m.s. amplitude of 8--10\%, FWHM of about 100 Hz, and smooth relation
of its frequency to that of QPOs and noise features below 100 Hz. No
systematic differences in its other properties were found when this
QPO varied in frequency between 725 Hz and 320 Hz. }
\end{center}
\end{figure}

\begin{figure}[t]
\begin{center}
\begin{tabular}{c}
\psfig{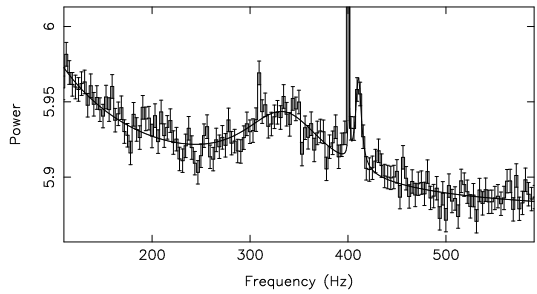}
\end{tabular}
\figcaption{\label{fig:f2}\small
The power-density spectrum obtained for SAX J1808.4--3658 using the
combined data from 18--19 October and 22--26 October. The power is
given in units of (r.m.s./mean)$^2$/(Hz $\times 10^{-3}$). This
spectrum is for the photon energy range 3--60 keV, and shows the
presence of the pulsations at 401 Hz, the upper kHz QPO, the
$\sim$410-Hz QPO, and broadband noise below a few hundred hertz. The
spectrum was fitted with the same function as described in Fig. 1, but
with an extra lorentzian function to fit the extra QPO at $\sim$410
Hz. The extra QPO could be detected significantly on four occasions:
in the combined 18--19 October data, the 22 October data, the combined
23--24 October data, and the combined 25--26 October data. Combining
all those data, the QPO was significant at a 6.5$\sigma$ level. This
QPO had an r.m.s amplitude of $\sim$2.5\% (for the energy range 3--60
keV) and its width (FWHM) was between 2 and 8 Hz.}
\end{center}
\end{figure}

\end{document}